\documentclass[
 reprint,
 preprintnumbers,
 amsmath,
 amssymb,
 aps,
 prstab,
 floatfix,
]{revtex4-1}

\usepackage{graphicx}			
\usepackage{verbatim}			
\usepackage{wasysym}			
\usepackage{mathtools}			
\usepackage[makeroom]{cancel}		
\usepackage{dcolumn}			
\usepackage{hyperref}			
\usepackage{color}			

\newcommand{\Nb}{N_{\text{p}}}		
\newcommand{\Qb}{ Q_{\beta}}		
\newcommand{\Qs}{ Q_s}		

\begin{document}

\title{TMCI with Resonator Wakes}
\author{A.~Burov}
\email{burov@fnal.gov}
\affiliation{Fermilab, PO Box 500, Batavia, IL 60510-5011}
\author{T.~Zolkin}
\affiliation{Fermilab, PO Box 500, Batavia, IL 60510-5011}
\date{\today}

\begin{abstract}
Transverse mode-coupling instability (TMCI) with a high-frequency resonator wake 
is examined by the Nested Head-Tail Vlasov solver (NHT), where a Gaussian bunch 
in a parabolic potential (GP model) is represented by concentric rings in the 
longitudinal phase space. It is shown that multiple mode couplings and 
decouplings make impossible an unambiguous definition of the threshold, unless 
Landau damping is taken into account. To address this problem, instead of a 
single instability threshold, an interval of thresholds is suggested, bounded by 
the low and high intensity ones. For the broadband impedance model, the high 
intensity threshold is shown to follow Zotter's scaling, but smaller by about a 
factor of two. The same scaling, this time smaller than Zotter's by a factor of 
four, is found for the ABS model (Air Bag Square well).     
\end{abstract}


\maketitle

\section{Introduction}

In principle, there are two distinct approaches to beam stability problems: the 
beam can be represented by a set of either macroparticles or continuous objects. 
The first choice leads to macroparticle tracking simulations, while the second 
one is applied in Vlasov solvers. Each of the methods has its advantages and 
drawbacks. The main advantage of the macroparticle approach lies in its 
potential to be closer to reality. This closeness may require, though, a lot of 
macroparticles, i.e. a lot of CPU time, which is a serious drawback. Vlasov 
solvers can be many orders of magnitude faster than typical macroparticle 
programs, but it remains to be unresolved how to include there, with accuracy 
and effectiveness, such factors as the space charge, certain kinds of Landau 
damping, or the electron cloud. 

This paper deals with a problem where Vlasov solvers apparently demonstrate 
their power in full: the transverse mode coupling instability (TMCI) without 
space charge. This instability used to be considered as one with a sharp 
threshold, i.e. which growth rate rapidly increases with the bunch intensity 
above its certain threshold value. That sort of TMCI behavior was demonstrated 
for the non-resonant wake functions, like the resistive wall one, possibly 
accompanied by a damper, see ref. \cite{PhysRevSTAB.17.021007, 
PhysRevAccelBeams.19.084402}. If the calculated threshold is sharp, even a 
strong variation of the Landau damping may only slightly change the onset of the 
instability, thus justifying the omission of the Landau damping in the threshold 
computation. A question may be asked though: is the TMCI threshold always as 
sharp as for the resistive wall type wakes? This question is especially 
motivated for short or highly oscillating wakes within the bunch length, where 
multiple mode couplings and decouplings may be important. Being driven by these 
questions, we limit our consideration here by high-frequency resonator wakes, 
i.e. by those which wavelength is shorter than the bunch length. 

Using NHT Vlasov solver \cite{PhysRevSTAB.17.021007}, it is shown below that the 
TMCI threshold is not actually sharp for the high-frequency wakes: as the bunch 
intensity begins to increase, the first couplings that appear are weak, being 
soon followed by the decouplings. With further increase of the intensity, mode 
couplings become stronger, with higher growth rates and larger 
coupling-decoupling intervals. As a result, the growth rate rises rather 
gradually with the intensity, making the very concept of the {\it{TMCI 
threshold}} vague unless the Landau damping is specified. Such sort of 
instability, with a cascade of the thresholds, was observed at the SPS, see e.g. 
Refs.~\cite{Salvant:2010dda, Bartosik:2013qji}. 

To deal with this conceptual uncertainty, the TMCI threshold can be represented 
by two intensities, low and high, with the former showing the first, very weak, 
mode coupling, and the latter pointing to a case where the growth rate increases 
sharply, soon after its appearance becoming comparable with the synchrotron 
tune. The asymptotical scaling of the high-intensity threshold is found in 
agreement with estimations of B.~Zotter \cite{Zotter:1982eb} and E.~Metral 
\cite{Metral:2002sw}, with a difference of numerical coefficients by a factor of 
$\sim 2$.   

A sensitivity of the TMCI thresholds to the shape of the potential well was 
examined by comparison of the Gaussian bunch in the parabolic bucket with the 
air-bag bunch in the square well. For the latter, we used Blaskiewicz' model as 
implemented in Ref. \cite{Zolkin:2017sdv}. Reasonable agreement of the two 
models is demonstrated.



\section{Parabolic potential well}

The resonator wake function can be presented as 
\begin{equation}
	W(\tau) = W_0 \sin(\widehat\omega\,\tau)\,
	e^{\omega_r\tau/(2\,Q_r)}; \;\;	
	W_0=R_s\,\frac{\omega_r^2}{\widehat\omega\,Q_r}\,,
\end{equation}
where $R_s$ is the transverse shunt impedance, frequency of oscillation
\begin{equation}
	\widehat\omega = \sqrt{\omega_r^2-\alpha^2},
\end{equation}
and the decay rate related to the resonant frequency $\omega_r$ by means of the 
Q-factor $Q_r$:
\begin{equation}
	\alpha = \omega_r/(2\,Q_r).
\end{equation}

At zero chromaticity, assumed at this paper, the problem is expressed in terms 
of the following dimensionless parameters: 
\begin{equation}
\chi = \frac{\Nb r_0 R_0 W_0}{4\,\pi\,\gamma\,\beta^2 \Qb \Qs}=\frac{\Nb 
r_0}{\gamma\,\beta\,\Qb \omega_s} \frac{R_s \omega_r^2}{\widehat\omega\,Z_0 Q_r}
\label{chi}
\end{equation}
\begin{equation}
\psi_r = \omega_r \sigma_{\tau} \equiv 2\,\pi\,\nu_r ;\qquad
\; \Delta_r = \alpha\,
\sigma_{\tau}. 
\label{psi}
\end{equation}
Here $Z_0=4\,\pi/c=377$~Ohm, $r_0$ is the particle classical radius, $\Nb$ is 
the 
number of particles per bunch, $\sigma_{\tau}$ is the rms bunch length,
$R_0=C_0/(2\,\pi)$ is the average accelerator ring radius,
$\Qb$ is the betatron tune, $\Qs$ and $\omega_s$ are the synchrotron tune and 
angular frequency,
$\gamma$ and $\beta$ are the Lorentz factors. 

An important motivation for this paper was understanding of the instability at 
CERN SPS, where the threshold was both measured and computed. Assuming the high 
frequency resonator wake, the latter has been done with the MOSES Vlasov solver 
and the HEADTAIL tracking \cite{Salvant:2010dda, quatraro2010effects} for the 
old Q26 lattice, and recently repeated for the new Q20 optics in the tracking 
simulations \cite{Oeftiger2018}. A surprising result of all the computations, 
both old and new, was nearly complete independence of the threshold on the space 
charge. In order to understand that, a more detailed look at the instability 
without the space charge for the high frequency broadband impedance model is 
needed as the first step. To make it easier, we are presenting our results in 
terms of the threshold number of particles per bunch $\Nb$, assuming the SPS 
optics Q20 with the gamma transition $\gamma_t=18$, at injection energy 26~GeV, 
the synchrotron tune $Q_s=0.017$ and the rms bunch length $\sigma_s = c\,
\sigma_\tau=23$~cm. In terms of the dimensionless $\chi$ parameter, Eq. 
(\ref{chi}), the relation between that and the number of particles can be 
presented as: 
\begin{equation}
\chi = 10\,\frac{\Nb}{10^{11}} \frac{\omega_r}{2\,\pi \cdot 1.3\,\mathrm{GHz}} 
\frac{R_s}{7\,\mathrm{M\Omega/m}} \sqrt{\frac{3}{4\,Q_r^2-1}}.
\label{chiNb}
\end{equation}    

\begin{figure}[h!]
\includegraphics[width=1.0\linewidth]{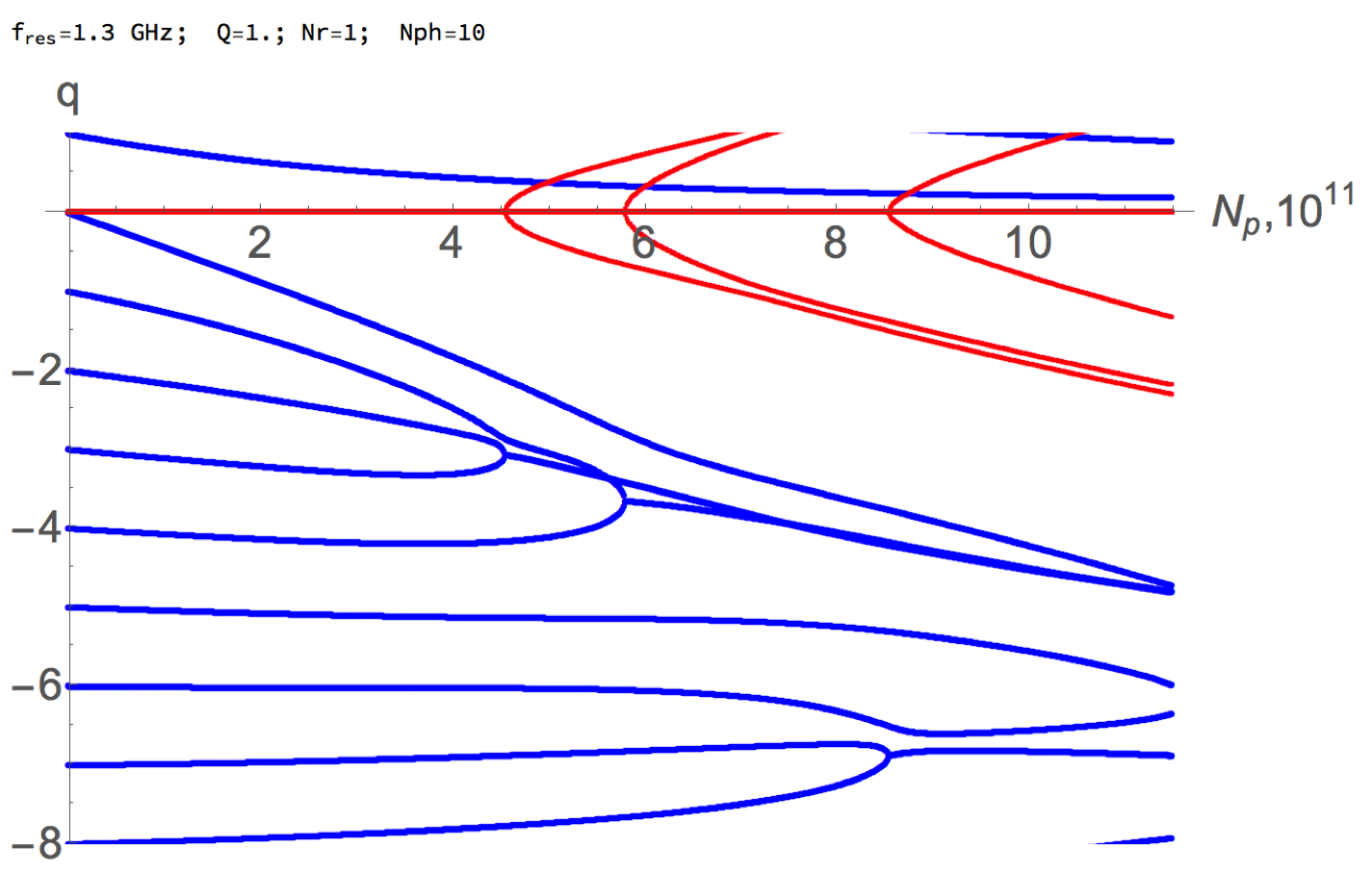}
\caption{\label{fig:NHT_1p3_Q1_Nr1_Nph10}
	Bunch modes versus the number of particles for the SPS at injection and 
Q20 optics. Coherent tunes $q$ are given in units of the synchrotron tune; real 
and imaginary parts are blue and red respectively. The resonator impedance is 
taken with $f_r=\omega_r/(2\,\pi)=1.3\,\mathrm{GHz}$, $Q_r=1$ and 
$R_s=7\,\mathrm{M\Omega/m}$. The bunch distribution is represented by a single 
ring in the longitudinal phase space, which is the air-bag model, with the 
radius $\sqrt{2}\,\sigma_s$. The head-tail harmonics are truncated at $\pm 10$. 
The resonant frequency, Q-factor, number of rings and head-tail truncation 
number are noted above the plot. 
		}
\end{figure}

Let's start from the SPS impedance model of the broadband resonance at 
$f_r=\omega_r/(2\,\pi)=1.3\,\mathrm{GHz}$, $Q_r=1$ and 
$R_s=7\,\mathrm{M\Omega/m}$, 
with the air-bag approximation. The bunch modes for various intensities are 
presented in Fig.~\ref{fig:NHT_1p3_Q1_Nr1_Nph10}.

The air-bag model works with variations of the dipole moment along the 
synchrotron phase; it does not take into account the variations along the 
synchrotron action, the so called radial harmonics. In the NHT Vlasov solver, 
any number of the radial rings can be taken to represent the bunch phase space 
distribution; thus, the radial harmonics can be included.  

\begin{figure}[h!]
\includegraphics[width=1.0\linewidth]{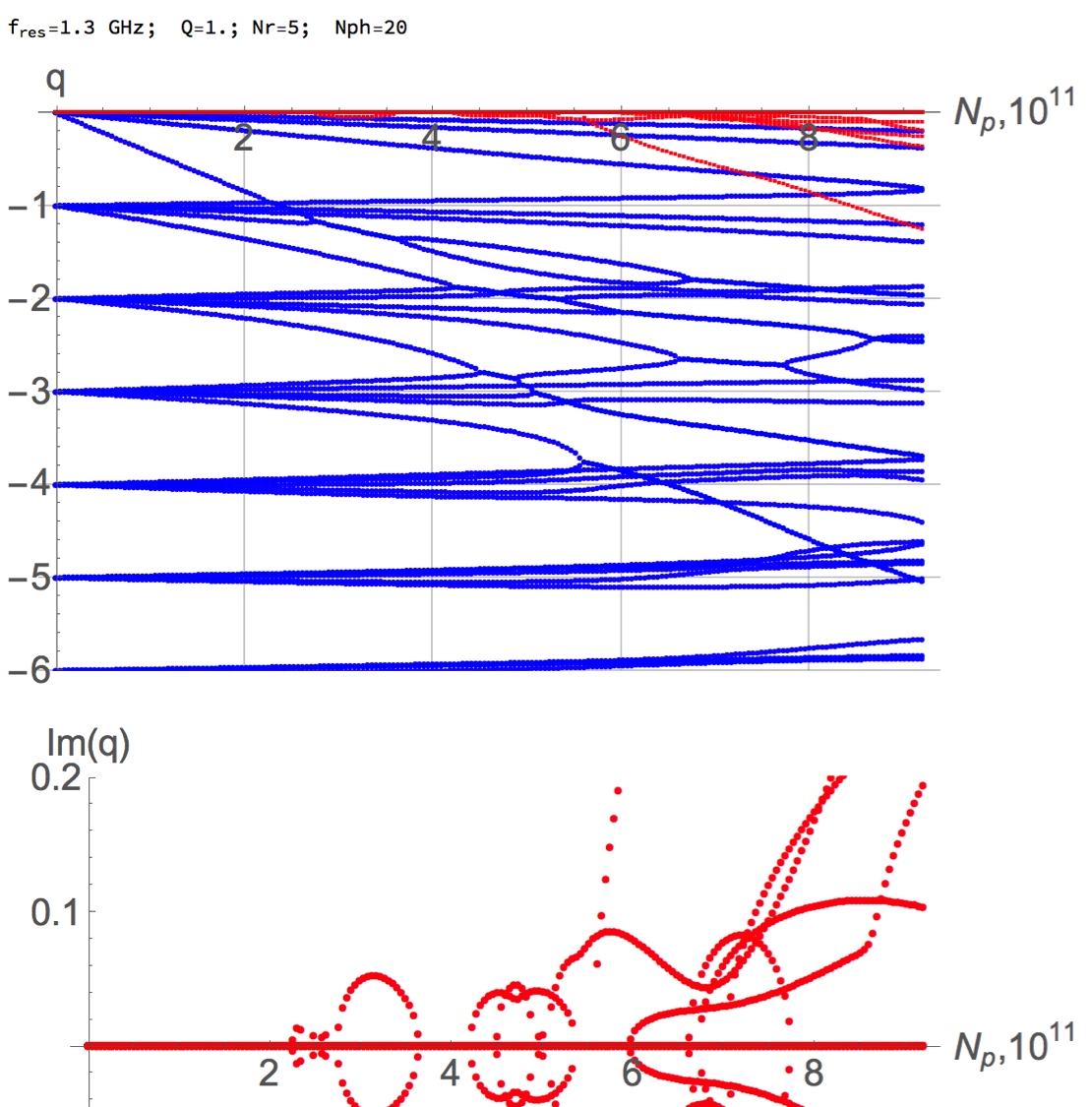}
\caption{\label{fig:FigNHT_1p3_Q1_Nr5_Nph20}
	Same as Fig. \ref{fig:NHT_1p3_Q1_Nr1_Nph10}, but with 5 NHT rings and 
head-tail truncation at $\pm 20$. 
		}
\end{figure}

\begin{figure}[h!]
\includegraphics[width=1.0\linewidth]{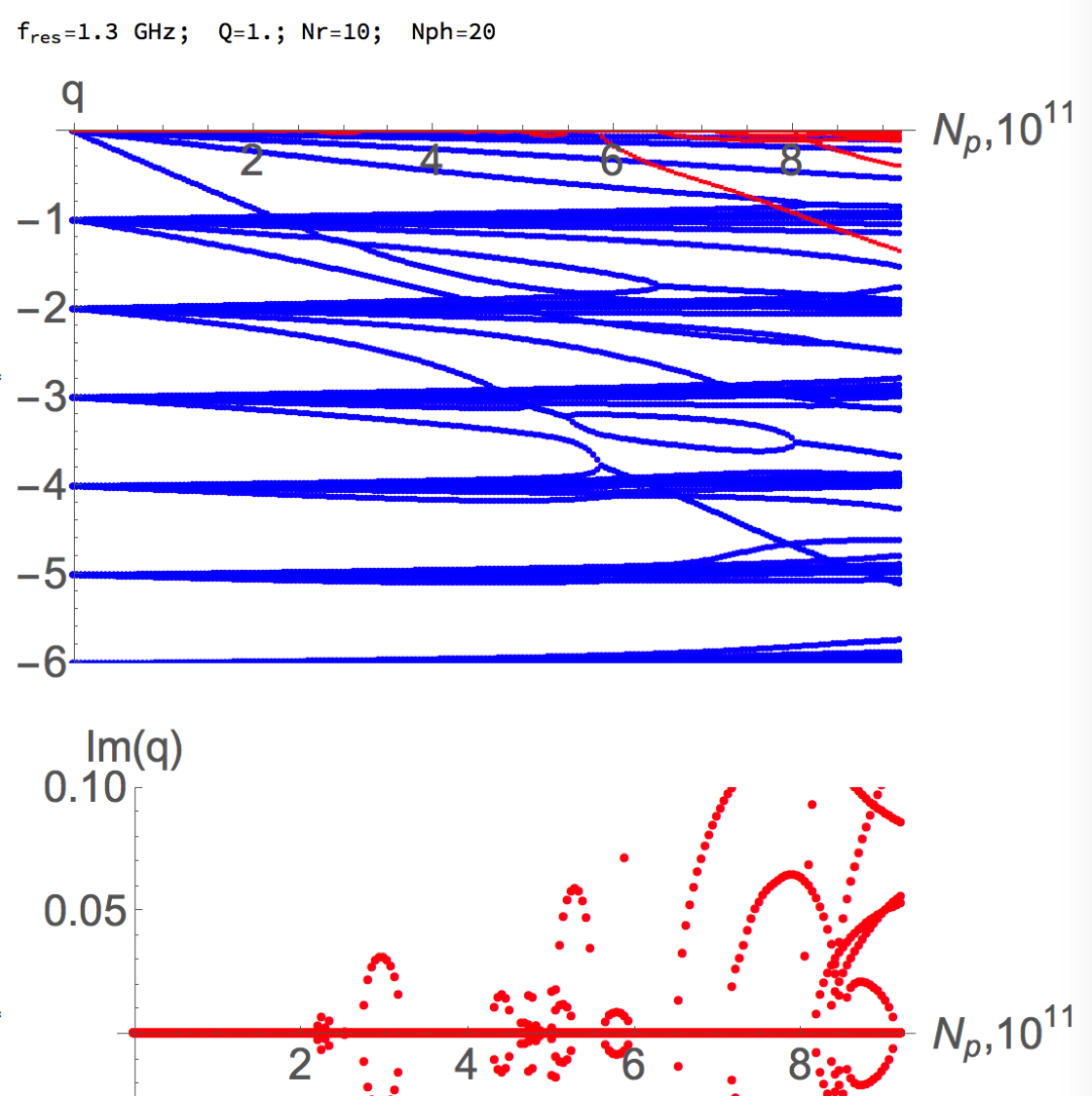}
\caption{\label{fig:FigNHT_1p3_Q1_Nr10_Nph20}
	Same, but with 10 NHT rings and head-tail truncation at $\pm 20$.
		}
\end{figure}

Figures \ref{fig:FigNHT_1p3_Q1_Nr5_Nph20} and \ref{fig:FigNHT_1p3_Q1_Nr10_Nph20} 
show how sensitive the instability is to the radial degrees of freedom. Both 
plots describe the same bunch and impedance as the air-bag, 
Fig.~\ref{fig:NHT_1p3_Q1_Nr1_Nph10}, but with more accuracy, taking 5 and 10 
radial rings respectfully, as well as double the azimuthal harmonics. Although 
there is some visible difference between Figs. \ref{fig:FigNHT_1p3_Q1_Nr5_Nph20} 
and \ref{fig:FigNHT_1p3_Q1_Nr10_Nph20}, their main features are similar, so we 
may reasonably consider them converged, and compare them together, as the 
many-ring case, the Gaussian bunch in the Parabolic bucket (GP model), with the 
air-bag case. Below we itemize some features of these plots. 
\begin{itemize}
\item For many rings the mode couplings begin at about factor of 2 lower 
intensity than for the air-bag (ABP) model. The first couplings are weak, soon 
followed by the decouplings and new, stronger, couplings. In this way, the 
instability growth rate goes up and down, increasing on the average. Similar 
mode behavior was recently observed with GALACTIC Vlasov solver for the 
broadband impedance with $\nu_r=0.7$ \cite{MetralIPAC18}. 
\item The intensity spread between the weak and strong couplings is about factor 
of 3 for these parameters. Without sufficient knowledge of the Landau damping, 
the real onset of the instability can be predicted only with that poor 
accuracy. 
\item Coupling of the negative 2nd and 3rd azimuthal harmonics happens at the 
intensity of $\Nb=4.5 \cdot 10^{11}$, in agreement with the pyHEADTAIL 
macroparticle tracking simulations \cite{Metral2017SCworkshop}. 
Figure~\ref{fig:FigNHT_1p3_Q1_Nr10_Nph20} suggests that for those simulations, 
the effective Landau damping rate was as small as $\leq 0.05\,\omega_s$.  
\end{itemize}

\begin{figure}[h!]
\includegraphics[width=1.0\linewidth]{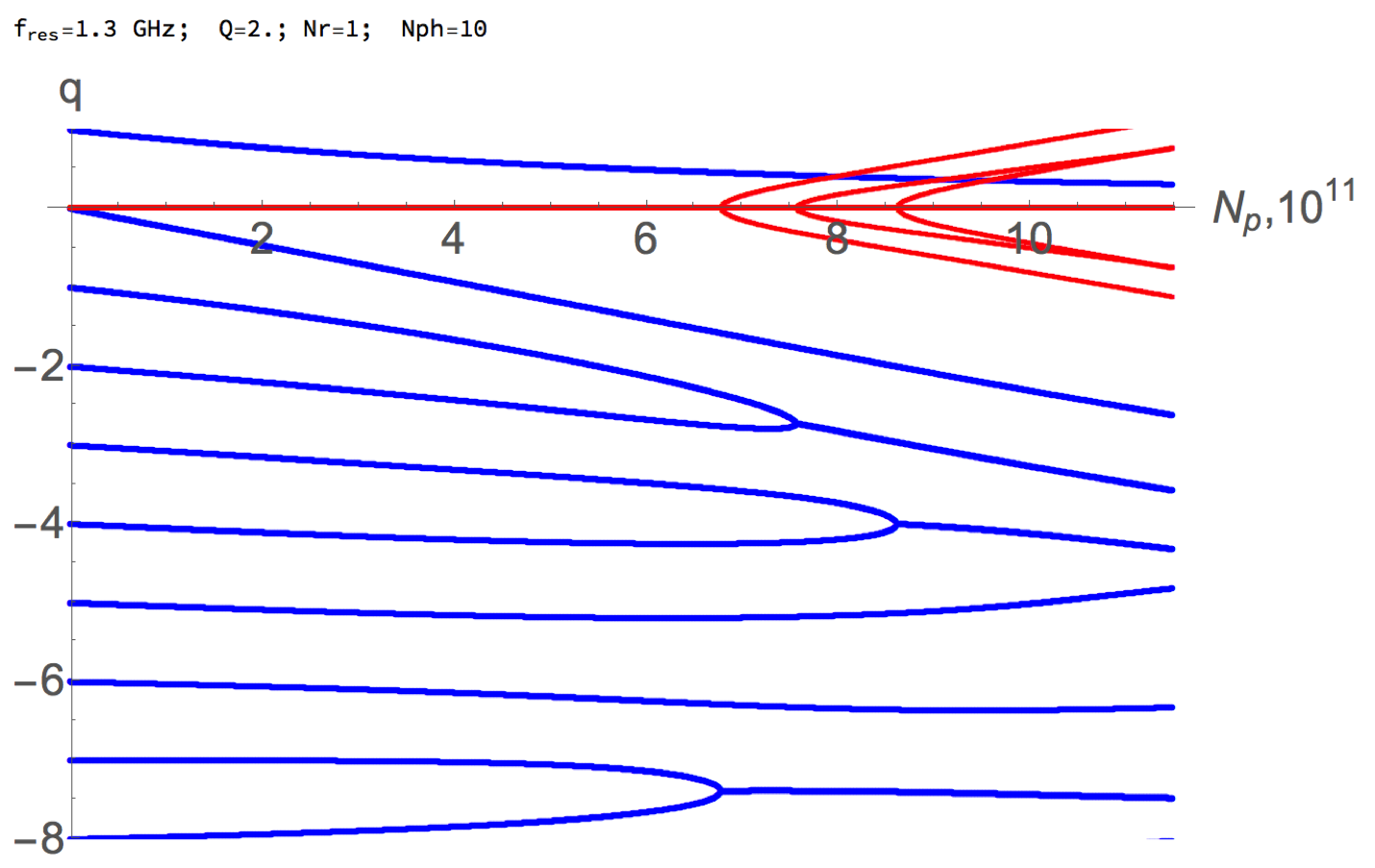}
\caption{\label{fig:FigNHT_1p3_Q2_Nr1_Nph10}
	Same shunt impedance and frequency, but $Q_r=2$. Air-bag, head-tail 
truncation at $\pm 10$. 
		}
\end{figure}

\begin{figure}[h!]
\includegraphics[width=1.0\linewidth]{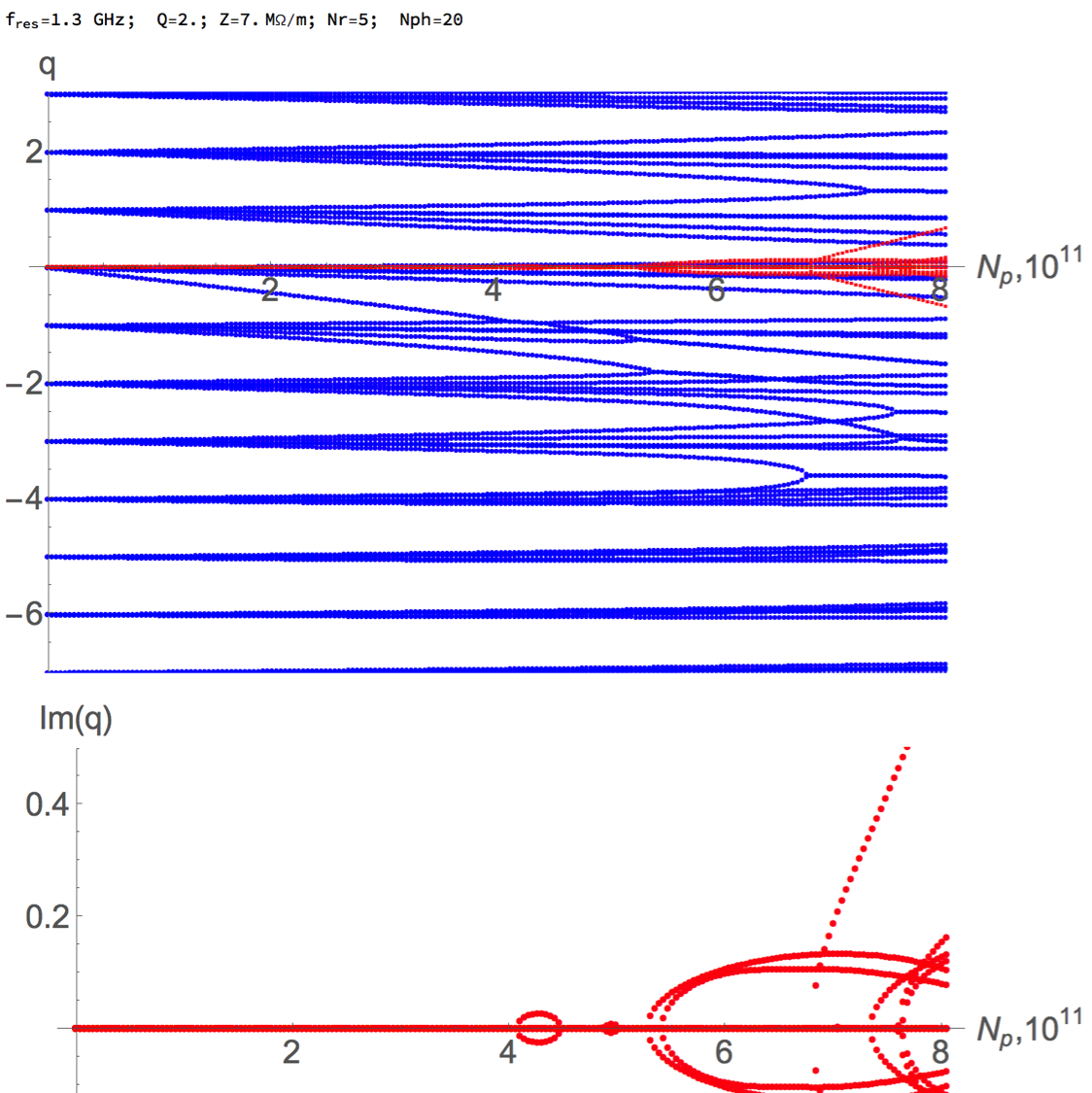}
\caption{\label{fig:FigNHT_1p3_Q2_Nr5_Nph20}
	Same, with 5 rings and truncation at $\pm 20$. Note that the positive 
modes may also couple, but this does not play a significant role. 
		}
\end{figure}

\begin{figure}[h!]
\includegraphics[width=1.0\linewidth]{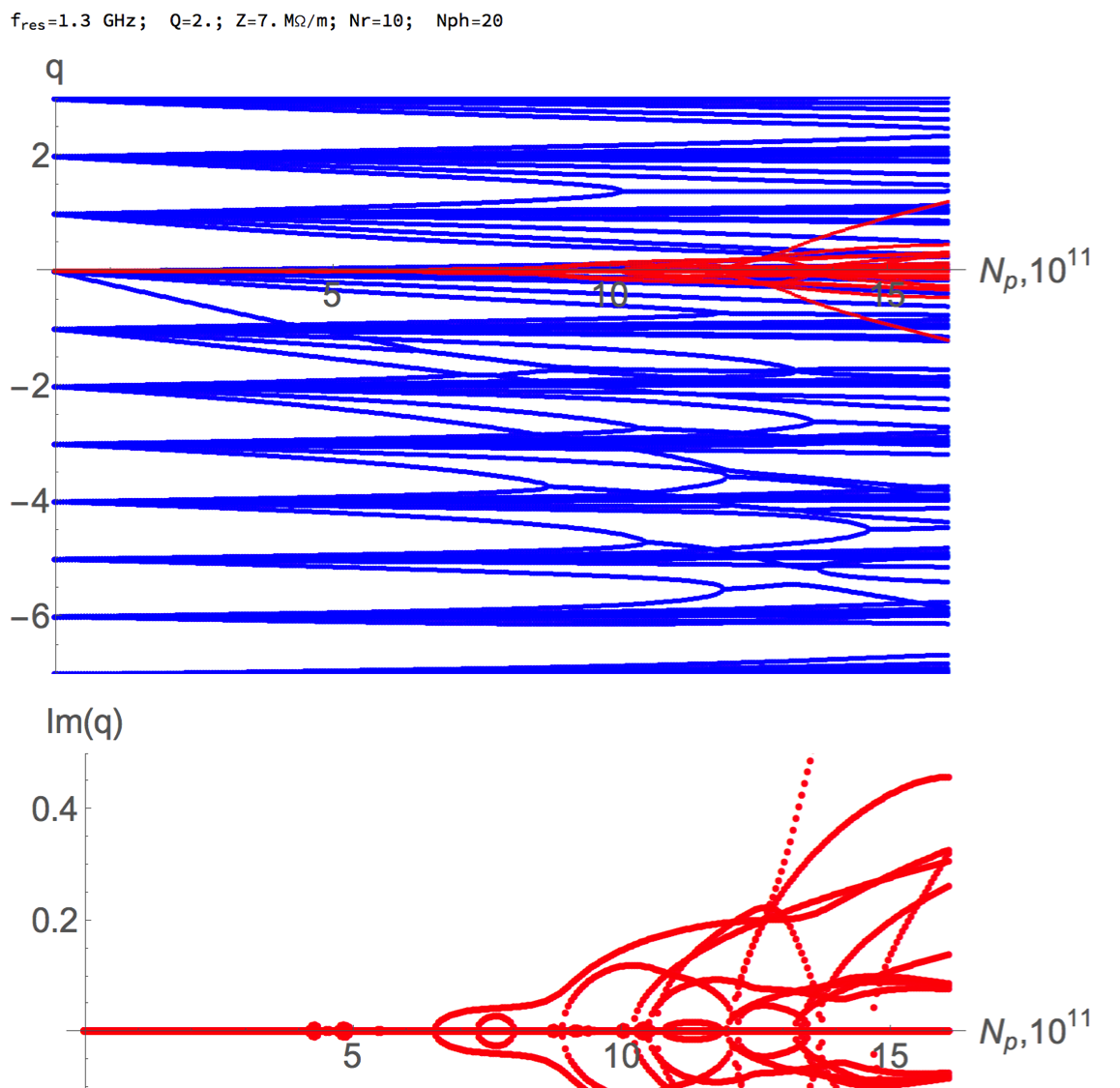}
\caption{\label{fig:FigNHT_1p3_Q2_Nr10_Nph20}
	Same, with 10 rings. 
		}
\end{figure}

How does the instability depend on the wake Q-value, $Q_r$? Some information on 
that is suggested by Figs. \ref{fig:FigNHT_1p3_Q2_Nr1_Nph10}, 
\ref{fig:FigNHT_1p3_Q2_Nr5_Nph20} and  \ref{fig:FigNHT_1p3_Q2_Nr10_Nph20}, which 
show the modes computed for the same shunt impedance and resonator frequency, 
but doubled $Q_r$. Comparing these plots with the previous three, we may 
conclude that in this case the thresholds increased twofold, as the Q-value. It 
is worth to note that with higher $Q_r$, convergence requires more rings.  


\begin{figure}[h!]
\includegraphics[width=1.0\linewidth]{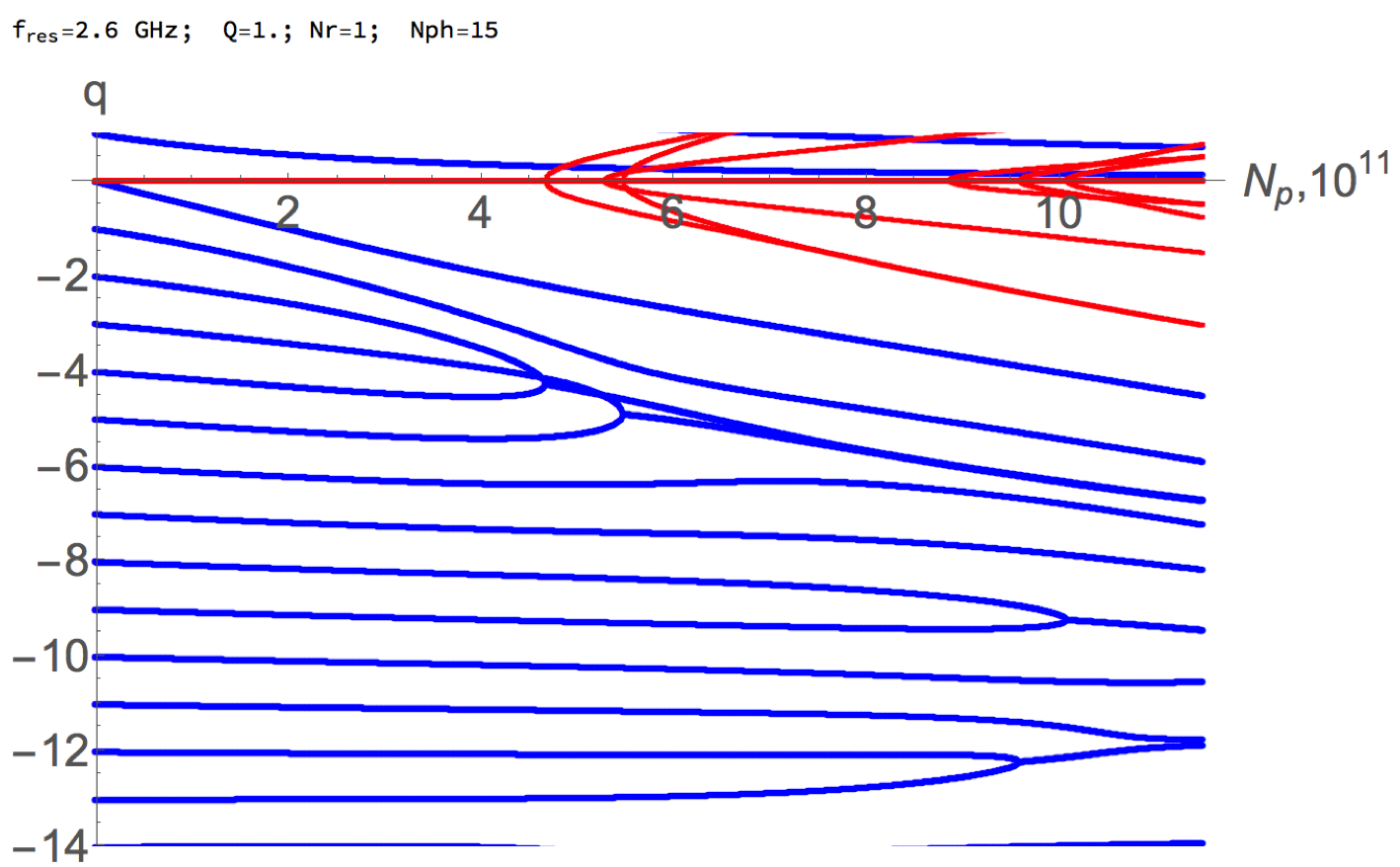}
\caption{\label{fig:FigNHT_2p6_Q1_Nr1_Nph15}
	The resonator frequency is doubled, $f_r=2.6$~GHz, the shunt impedance 
is 
the same, $Q_r=1$, air-bag, head-tail truncation at $\pm 15$.
		}
\end{figure}

\begin{figure}[h!]
\includegraphics[width=1.0\linewidth]{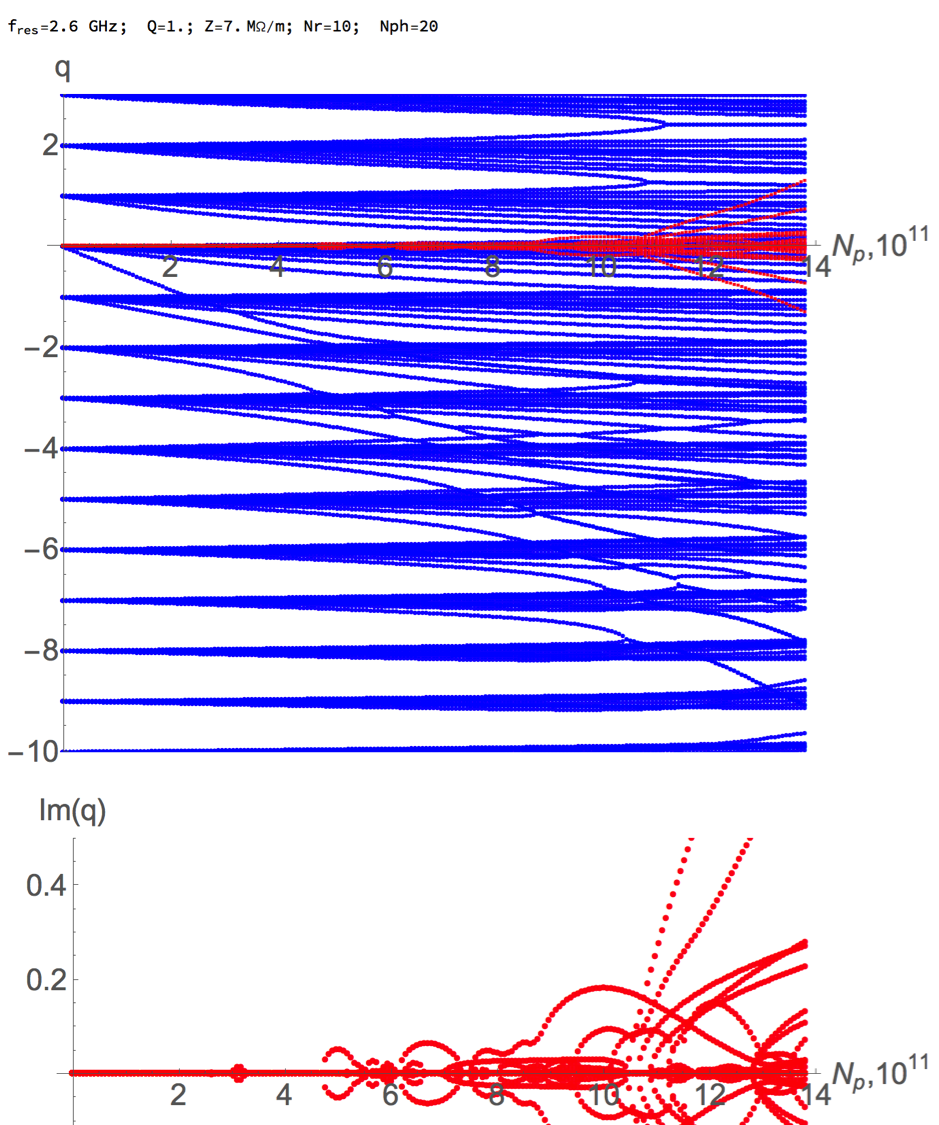}
\caption{\label{fig:FigNHT_2p6_Q1_Nr10_Nph20}
	Same as the previous Figure, but with 10 rings and the truncation at 
$\pm 20$. 
		}
\end{figure}

Figs. \ref{fig:FigNHT_2p6_Q1_Nr1_Nph15}, \ref{fig:FigNHT_2p6_Q1_Nr10_Nph20} 
illustrate how the instability depends on the resonator frequency $\omega_r$. 
Comparing them with the corresponding plots for half the resonator frequency, we 
may conclude that while the air-bag and low-intensity thresholds are almost the 
same, the high-intensity threshold for the multi-ring GP model is doubled for 
double the frequency.    


\begin{figure}[h!]
\includegraphics[width=1.0\linewidth]{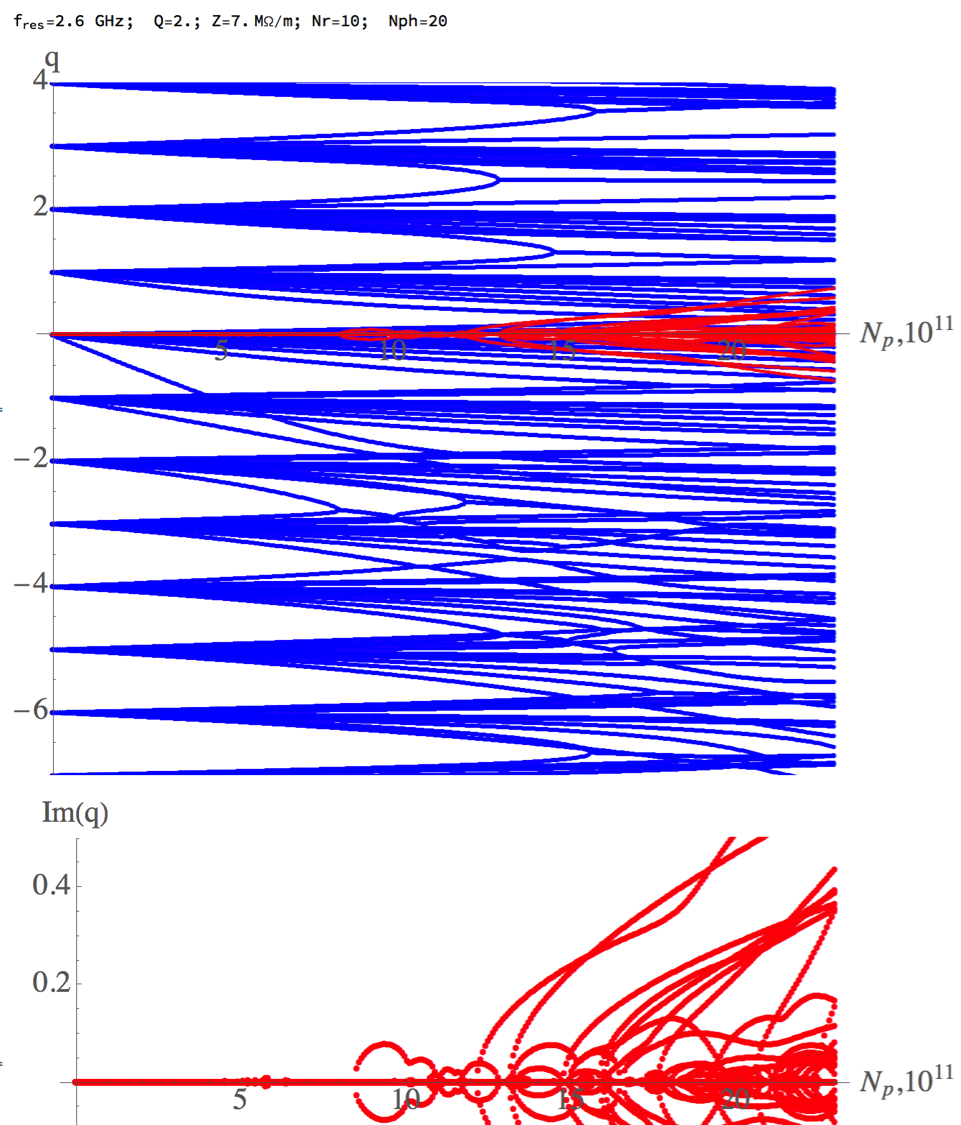}
\caption{\label{fig:FigNHT_2p6_Q2_Nr10_Nph20}
	$f_r=2.6$~GHz, $Q_r=2$.		}
\end{figure}

Looking at Figs.~\ref{fig:FigNHT_1p3_Q1_Nr10_Nph20}, 
\ref{fig:FigNHT_1p3_Q2_Nr10_Nph20}, one may guess that the thresholds will show 
proportionality to the Q-value for higher resonator frequency as well. However, 
comparison of Fig.~\ref{fig:FigNHT_2p6_Q2_Nr10_Nph20} with 
Fig.~\ref{fig:FigNHT_2p6_Q1_Nr10_Nph20}, show that it is not necessarily so: the 
two plots demonstrate almost the same high-intensity threshold, while their 
Q-values differ by the factor of two.  

The presented plots convince the authors that there is no good analytical 
fitting for the TMCI threshold versus frequency and Q-value for high-frequency 
resonator wake, $\omega_r \sigma_\tau \gg 1$, for the considered range of 
parameters. After all, there is no reason for that: in this sort of computation, 
finding the minimum of several functional branches is involved; taking the 
minimum is not an analytical procedure. 

What about the air-bag approximation? Can it be considered a reasonable way to 
estimate the instability threshold? 
Plots of the air-bag thresholds versus the wake phase advance $\nu_r=f_r 
\sigma_{\tau}$ for various Q-values are presented in Fig.~\ref{fig:FigChiABP}. 
They demonstrate that for $Q_r=1$, the fitting formula
\begin{equation}
\chi^{\mathrm{ABP}}_{\mathrm{th}}=40\,\nu_r
\label{chithABP}
\end{equation} 
works well; for all Q-values this linear dependence works better than other 
powers of $\nu_r$. In terms of the number of particles, it is equivalent to the 
threshold $N_\mathrm{th}$ being independent of the wake frequency $f_r$ when the 
shunt impedance $R_s$ is given; for $Q_r=1$:
\begin{equation}
N_\mathrm{th}^{\mathrm{ABP}}=6\,\frac{\gamma\,\beta\,\Qb \omega_s 
\sigma_\tau}{r_0} 
\frac{Z_0}{R_s}.
\label{NthABP}
\end{equation} 

\begin{figure}[h!]
\includegraphics[width=1.0\linewidth]{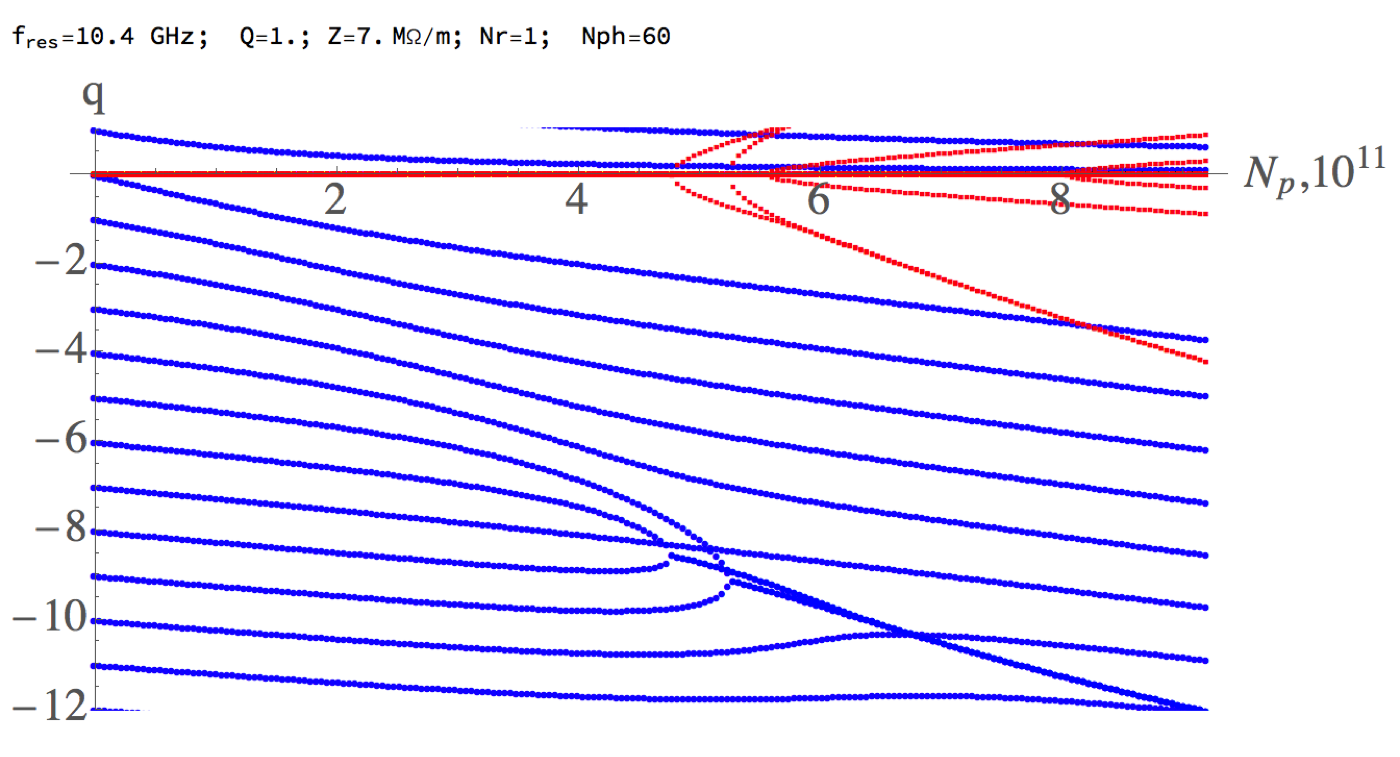}
\caption{\label{fig:FigNHT_10p4_Q1_Nr1_Nph60}
Air-bag with 10.4~GHz wake, same $R_s$ and $Q_r=1$. Although the threshold is 
same as for the lower frequency, the coupling mode numbers are higher.}
\end{figure}

\begin{figure}[h!]
\includegraphics[width=1.0\linewidth]{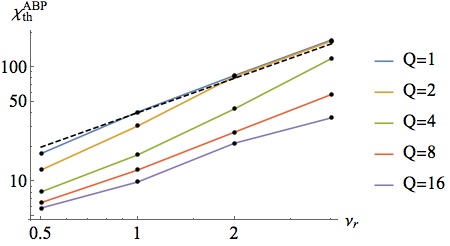}
\caption{\label{fig:FigChiABP}
Threshold $\chi$ parameters for the ABP model versus the phase advance 
$\nu_r=f_r \sigma_{\tau}$ at various Q-values. The dashed line is a linear fit 
$\chi^{\mathrm{ABP}}_{\mathrm{th}}=40\,\nu_r$ for $Q_r=1$. }
\end{figure}

Such dependence means that only an integral of the wake, i.e. the shunt 
impedance, is important for the threshold number of particles. Figure 
\ref{fig:FigNHT_10p4_Q1_Nr1_Nph60} demonstrates that for even an higher 
frequency of $10.4$~GHz, the threshold is still the same, albeit the numbers of 
the coupling modes are larger. This leads us to the conclusion that the same 
threshold has to be expected for the delta-function wake, $W(\tau) =R_s 
\delta(\tau)$. On another hand, it can be proven that the delta-wake cannot lead 
to any instability of a thermalized bunch, independently of the potential well. 
Indeed, the delta-wake is equivalent to a combination of the space charge and 
longitudinally-dependent external transverse focusing, so the system can be 
described by a time-independent Hamiltonian. Such systems in thermal equilibrium 
must be stable: the opposite would violate the Second Law of Thermodynamics. 
Thus, the instability threshold of Eq. (\ref{NthABP}) may follow only from the 
inadequacy of the ABP model for the Gaussian bunch, where such dependence 
shouldn't be possible. The phase space density of the air-bag approximation is a 
highly non-monotonic function of the longitudinal action; moreover, at the bunch 
edges, $\tau=\pm \tau_b/2 \equiv {\hat{\tau}}$, the ABP line density has 
singularities $\sim 1/\sqrt{{\hat{\tau}}^2 -\tau^2}$. Those unphysical 
properties of the ABP model may be reasonably tolerable for long-range wake 
functions $\nu_r \ll 1$, but they lead to inadequate results for the short-range 
ones, $\nu_r \geq 1$, when the role of the line density singularities is more 
pronounced. This conclusion is supported by comparison of 
Figs.~\ref{fig:FigNHT_1p3_Q1_Nr10_Nph20} and \ref{fig:FigNHT_2p6_Q1_Nr10_Nph20}, 
taken for doubled wake frequencies and computed with the sufficient number of 
the rings. While the air-bag model gives the same threshold number of particles 
for the two cases, the multi-ring GP model points to the increase of both low 
and high intensity thresholds. The most interesting of them, the latter, 
increases by about a factor of 2, proportionally to the frequency. Led by this 
observation, one may suggest the following high-intensity threshold for $Q_r=1$: 
          
\begin{equation}
\chi_h=60\,\nu_r^2.
\label{ChithParHigh}
\end{equation} 
or
\begin{equation}
N_h=1.5\,\frac{\gamma\,\beta\,\Qb \omega_s \omega_r \sigma_\tau^2}{r_0} 
\frac{Z_0}{R_s},
\label{NthParHigh}
\end{equation} 
where the subscript $h$ stands for {\it{high}}. 

The frequency scaling of Eq.(\ref{NthParHigh}) agrees with one predicted by 
B.~Zotter \cite{Zotter:1982eb} and E.~Metral \cite{Metral:2002sw}, while their 
numerical factors are about two times larger. On another hand, as shown in the 
next section, the high intensity GP instability threshold (\ref{NthParHigh}) 
exceeds by factor of 2 the high-frequency asymptote of the TMCI threshold for 
the ABS model. Elias Metral drew our attention to an article of J.-L. 
Laclare~\cite{laclare1987bunched}, where the TMCI threshold for high-frequency 
broadband impedance was independently calculated. It was proposed by Metral, 
that there is a typo in the vertical axis label of the last Laclare's plot, 
Fig.~(37): instead of $\varepsilon_{th} Z_{\perp} p_r B_0$ there must be just 
$\varepsilon_{th} Z_{\perp}$. Indeed, the numbers in the textual surrounding of 
this plot confirm this guess very well. A possible reason for that sort of typo 
might be just lost round brackets, closing $p_r B_0$ as the argument. Following 
this guess, one may express the linear high-frequency dependence of Laclare's 
Fig.~(37) in terms of this paper, and obtain our Eq.~(\ref{NthParHigh}) with 
just slightly higher numerical factor, 1.7, instead of our 1.5.      

It's interesting that Zotter's scaling (\ref{NthParHigh}) is identical to one of 
the coasting beam, as it was noted by E.~Metral; this becomes obvious after a 
substitution $\omega_s \sigma_\tau^2 = |\eta|\,\epsilon_z/(\gamma\,\beta^2 
m\,c^2) 
$, where $\eta$ is the slippage factor and $\epsilon_z$ is the rms longitudinal 
emittance. If this agreement is not a coincidence, the same scaling would be 
valid rather independently of the Q-value. Our current data and understanding of 
the issue does not allow us making a definite statement on that.    

\begin{figure}[h!]
\includegraphics[width=1.0\linewidth]{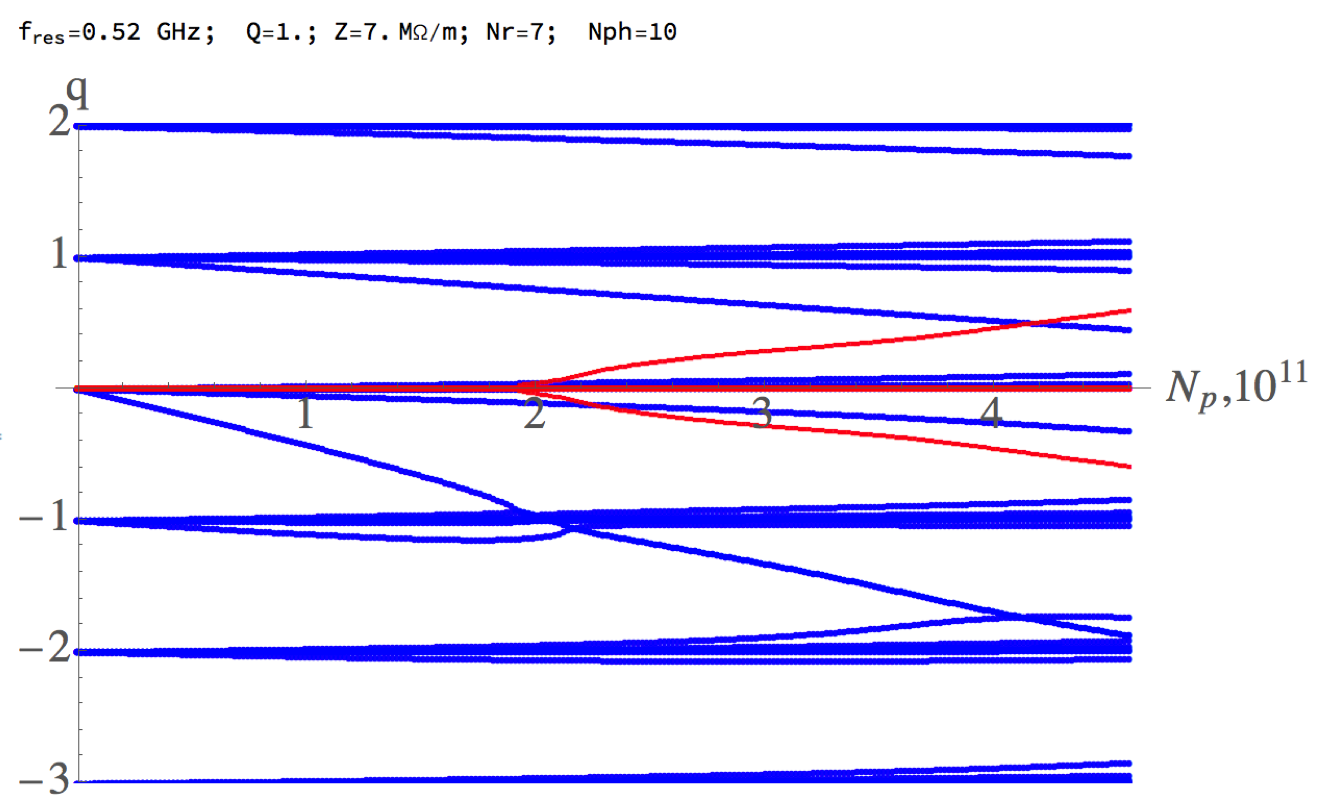}
\caption{\label{fig:FigNHT_0p52_Q1_Nr7_Nph10}
The spectrum at the resonator frequency just slightly above the high-low merge, 
$\nu_r=0.4$. Note the tiny intensity interval between a week and a strong 
coupling of the mode $0$ with different modes $-1$. Below this frequency, the 
TMCI threshold is sharp.}
\end{figure}

The low-intensity threshold can be fitted as
\begin{equation}
\chi_l=20\,\nu_r; \;\; N_l=3\,\frac{\gamma\,\beta\,\Qb \omega_s 
\sigma_\tau}{r_0} 
\frac{Z_0 Q_r}{R_s},
\label{ChiNthParLow}
\end{equation} 
with the subscript $l$ standing for {\it{low}}. For $Q_r=1$, the two thresholds 
merges at $\omega_r \sigma_\tau \approx 2$.  For the SPS Q20 parameters it 
corresponds to $f_r=0.45$~GHz. According to that, the TMCI threshold must be 
sharp at frequency below this value. Figure~\ref{fig:FigNHT_0p52_Q1_Nr7_Nph10} 
confirms that, showing the spectra slightly above this high-low merge, when the 
intensity interval between them is almost zero. Although the low-intensity 
threshold (\ref{ChiNthParLow}) has the same scaling as the ABP one 
(\ref{NthABP}), the former does not necessarily contradict to the Second Law of 
Thermodynamics in the sense as the latter does. Indeed, with increase of the 
wake frequency, the instability growth rate at the low-intensity edge may become 
smaller and smaller, which tendency is confirmed by the presented plots.     


\section{Square potential well}
 
How sensitive are the thresholds to the potential well shape? In order to shed 
light on that, we consider a bucket rather dissimilar to the parabolic one: the 
square well. The analysis in the previous section forced us to conclude that the 
air-bag model is not a good representative of a Gaussian-like bunch in the 
parabolic potential well. However, this conclusion is not necessarily valid for 
the square well, where the line density never goes to infinity. The possibility 
of an exact analytical solution with an arbitrary space charge and any 
combination of the resonator wakes, demonstrated by M. Blaskiewicz 
\cite{blaskiewicz1998fast}, suggests that the air-bag bunch in the square well 
may be particularly interesting. Led by this interest, we consider here the 
instability for the air-bag distribution in a square well, the ABS model, 
following the same routine as we recently have \cite{Zolkin:2017sdv}. A space 
charge of zero is assumed here; instead, we vary the wake frequency and Q-value 
and study how the instability threshold changes with that. Our main task here is 
to see the level of agreement between the GP and the ABS models.

\begin{figure}[h!]
\includegraphics[width=1.0\linewidth]{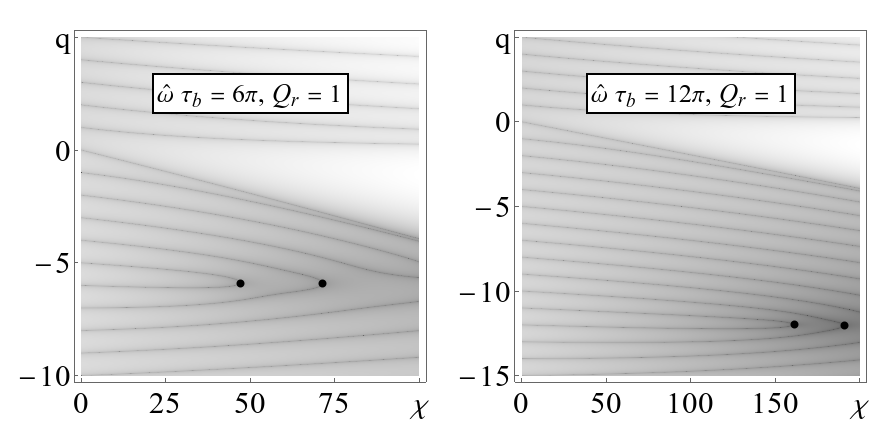}
\caption{\label{fig:SpectraABS_Omega6vs12pi}
Collective spectra for the ABS model. Only real tunes are shown; the mode 
couplings are marked by dots. Comparison of the left and right plots show that 
when the wake frequency doubles, numbers of the coupling modes double too.}
\end{figure}

For that, we take the total length of the ABS bunch, $\tau_b$, equal to three 
rms lengths of the GP case: $\tau_b=3\,\sigma_{\tau}$. With this convention, 
the 
SPS wake model with $f_r=1.3$~GHz results in the ABS phase advance $f_r \tau_b 
=3\,\nu_r=3$. Figure \ref{fig:SpectraABS_Omega6vs12pi} shows the coherent 
spectra 
for this and doubled resonator frequencies. Their comparison demonstrates that 
the numbers of the coupled modes grow linearly with the wake frequency, as $ 
2\,f_r \tau_b$. This is different from the GP case, where, as it was shown 
above, 
a series of couplings and decouplings were observed, where the instability rate 
increased on average with the bunch intensity.  

\begin{figure}[h!]
\includegraphics[width=1.0\linewidth]{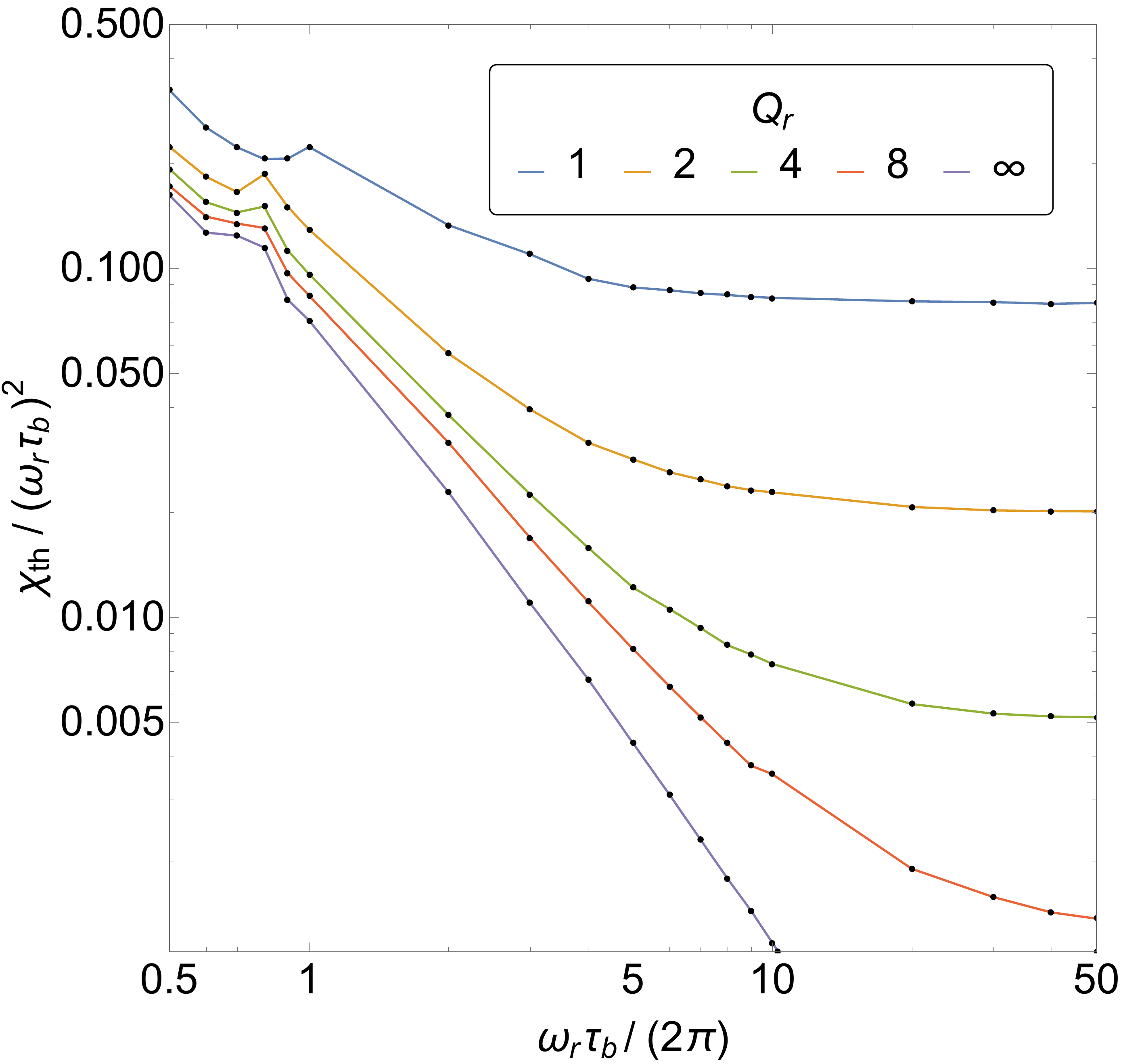}
\caption{\label{fig:Chi_ABS_ManyLines}
ABS thresholds versus phase advances for various Q-factors.}
\end{figure}

\begin{figure}[h!]
\includegraphics[width=1.0\linewidth]{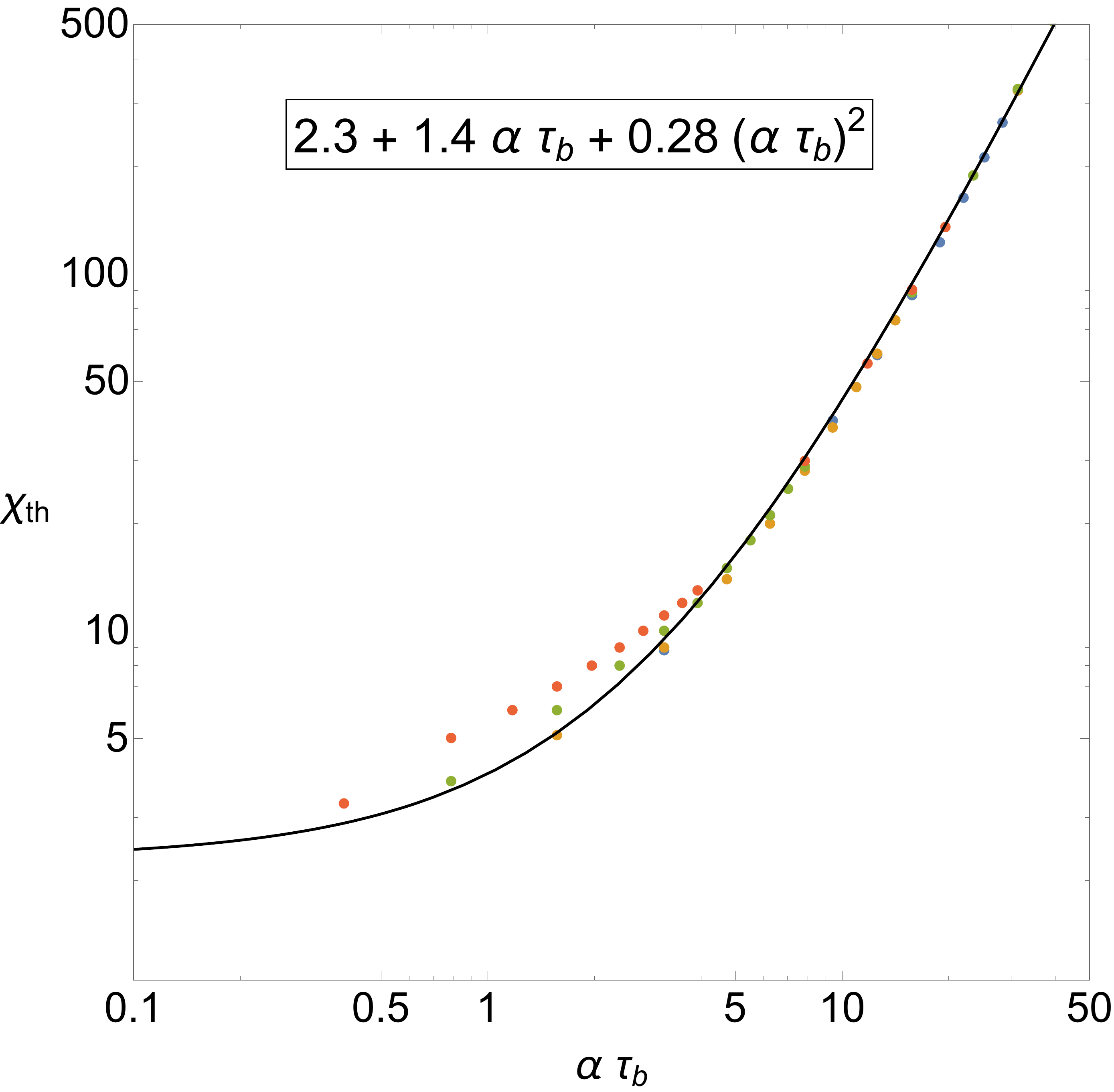}
\caption{\label{fig:Chi_ABS_Fit}
ABS thresholds can be reasonably fitted by the shown dependence on the wake 
decrement $\alpha\,\tau_b$. Color convention is the same as in the previous 
figure.}
\end{figure}

ABS thresholds versus phase advances are presented in Fig. 
\ref{fig:Chi_ABS_ManyLines}. The asymptotes suggest dependence $\propto 
1/Q_r^2$. This idea leads to a fitting of Fig. \ref{fig:Chi_ABS_Fit}, which 
appears to be rather good. At $Q_r=1$ and sufficiently high decrements, 
$\alpha\,
\tau_b \gg 3$, this dependence is of the same scaling as our GP result for the 
high-intensity threshold, being about half that result. For $Q_r=1$, the same 
scaling was suggested by B. Zotter \cite{Zotter:1982eb} and by E. Metral, Ref. 
\cite{Metral:2002sw}, with different numerical factors. Zotter's and Metral's 
thresholds are $\sim 3-4$ times larger than our ABS asymptotic value, 
correspondingly. This reasonable agreement of the ABS threshold with results of 
the more realistic GP model without any space charge adds more confidence to the 
ABS model when it takes the space charge into account, as we do in 
Ref.~\cite{Zolkin:2017sdv}.

For the accepted SPS impedance model with $f_r=1.3$~GHz, the rms bunch length 
$\sigma_{\tau} =0.77$~ns, and the phase advance $\nu_r=f_r \sigma_\tau=1.0$, 
the 
ABS model yields the threshold intensity $N_p = 4.5 \cdot 10^{11}$, which 
amazingly is almost identical to the one obtained by pyHEADTAIL tracking 
simulations, $N_p = 4.2 \cdot 10^{11}$ \cite{Metral2017SCworkshop, 
Oeftiger2018}. This number is about 1.5 times below the GP high-intensity 
threshold, suggesting a rather small Landau damping rate, $\leq 
0.05\,\omega_s$, 
for the tracking conditions. Similarly good agreement is found between the ABS 
and the HEADTAIL tracking simulations for the old Q26 optics 
\cite{quatraro2010effects}, giving the same threshold intensity $N_p=1.8 \cdot 
10^{11}$, although the coupling mode numbers are not the same.

\section{\label{sec:RnD}Summary}

Transverse mode coupling instability has been examined by means of the Nested 
Head-Tail Vlasov solver (NHT) for a Gaussian bunch in a parabolic bucket (GP 
model) and the ABS model (Air-Bag, Square well) for high-frequency resonator 
wakes, $f_r \sigma_\tau \geq 1$, with various phase advances $f_r \sigma_\tau $ 
and decrements $\alpha_r \sigma_\tau$. For the GP case, it has been found, in 
agreement with Ref.~\cite{Salvant:2010dda, Bartosik:2013qji}, that radial modes 
make the very concept of this threshold rather vague, due to multiple couplings 
and decouplings of various modes, with a gradually increasing growth rate. Thus, 
for high-frequency resonator wakes, the instability onset cannot be regularly 
predicted with good accuracy, unless Landau damping is taken into account. 
An appreciable agreement was seen between the NHT results and the HEADTAIL and 
pyHEADTAIL multiparticle simulations for the SPS broadband impedance model, 
\cite{quatraro2010effects, Metral2017SCworkshop, Oeftiger2018}: the same modes 
-2 and -3 couple at almost the same bunch intensities.     
It can be said that an amazingly good agreement was found between the ABS 
threshold and the tracking results: the thresholds do not differ by more than 
$\sim$10--15\%; however, the coupling mode numbers are not the same. A fit for 
the high-intensity threshold of the GP bunch with the broadband impedance, 
$Q_r=1$, was suggested, being $\sim 2$ times lower than the analytical 
predictions of B.~Zotter \cite{Zotter:1982eb} and E.~Metral 
\cite{Metral:2002sw}. The same high-frequency scaling law was obtained for the 
ABS model, with the numerical coefficient about half our high-intensity GP 
threshold.  
\begin{acknowledgments}

We are thankful to Adrian Oeftiger for explanations about multiple details of 
his pyHEADTAIL tracking simulations for the SPS instability. We express our 
special gratitude to Elias Metral for numerous informative and clarifying 
discussions.

This manuscript has been authored by Fermi Research Alliance, LLC under 
Contract 
No. DE-AC02-07CH11359 with the U.S. Department of Energy, Office of Science, 
Office of High Energy Physics. The U.S. Government retains and the publisher, 
by 
accepting the article for publication, acknowledges that the U.S. Government 
retains a non-exclusive, paid-up, irrevocable, world-wide license to publish or 
reproduce the published form of this manuscript, or allow others to do so, for 
U.S. Government purposes.
\end{acknowledgments}




\providecommand{\noopsort}[1]{}\providecommand{\singleletter}[1]{#1}%

\end{document}